\documentclass[a4paper,11pt]{article}
\pdfoutput=1 

\usepackage{jinstpub} 

\title{\boldmath Characterization and on-field performance of the MuTe Silicon Photomultipliers}

\author[a,2]{J. S\'anchez-Villafrades,}
\author[b,1]{J. Pe\~na-Rodr\'iguez,\note{Corresponding author.}}
\author[d,e,2]{H. Asorey}
\author[b,c,2]{and L. A. N\'u\~nez}

\affiliation[a]{Escuela de Ingenier\'ia El\'ectrica, Electr\'onica y de Telecomunicaciones, \\ Universidad Industrial de Santander, Bucaramanga-Colombia}
\affiliation[b]{Escuela de F\'isica, Universidad Industrial de Santander, Bucaramanga-Colombia}
\affiliation[c]{Departamento F\'isica M\'edica, Centro At\'omico Bariloche, Comisi\'on Nacional de Energ\'{\i}a At\'omica,\\ Bariloche-Argentina}
\affiliation[d]{Instituto de Tecnolog\'{\i}as en Detecci\'on y Astropart\'{\i}culas, Buenos Aires-Argentina.}
\affiliation[e]{Departamento de F\'isica, Universidad de Los Andes, M\'erida-Venezuela.} 

\emailAdd{jesus.pena@correo.uis.edu.co}

\abstract{The Muon Telescope is a muography experiment for imaging volcanoes in Colombia. It consists of a scintillator tracking system and a water Cherenkov detector used for particle deposited energy measurement. The MuTe operates autonomously in high altitude environments where the temperature gradient reaches up to $10$\,$^{\circ}$C. In this work, we characterize the breakdown voltage, gain, and noise of the telescope silicon photomultipliers for temperature variations spanning $0$ to $40$\,$^{\circ}$C. We demonstrated that the discrimination threshold for the MuTe hodoscope must be above $5$\, pe to avoid contamination due to the SiPM dark count, crosstalk, and afterpulsing. We also assess the MuTe counting rate depending on day-night temperature variations.}


\arxivnumber{xxxx.xxxxx} 

\begin{document}
\maketitle
\flushbottom

\section{Introduction}
\label{sec:intro}

Muography is a non-invasive technique for imaging anthropic and geologic structures \cite{Blanpied2015, Morishima2017, GomezEtal2016, Fujii2013, Saracino2017, ThompsonEtal2019, Tanaka2005, Tanaka2009, Lesparre2010, Lesparre2011, Lesparre2012, TanakaOlah2019} by measuring the crossing muon flux using sensitive hodoscopes made of nuclear emulsions \cite{Morishima2017, NAGAMINE2016}, gaseous chambers \cite{Sehgal2016, Fehr2012, Bouteille2016, Olh2018} and scintillators \cite{Fujii2013, Lesparre2012, Tanaka2009, Nagamine1995, Aguiar2015, Tang2016}. Scintillation hodoscopes provide flexibility on the implementation, low cost, and robustness against environmental variables such as humidity, temperature, and atmospheric pressure \cite{Procureur2018}. When an ionizing particle interacts with the scintillator crystal lattice, it knocks electrons out from the valence band to bound states called excitons. Then excitons emit photons in the near-ultraviolet spectrum because of de-excitation by recombination. Some dopants are added to the basic scintillation material to obtain a light output in a longer wavelength taking into account the absorption length of the ultraviolet light is quite short. The resultant emission range of the scintillator mismatches the sensitivity of the most photosensors (Photomultipliers or SiPMs) being necessary to add a wavelength shifting fiber \cite{Grupen2008}. 


SiPMs offer a solution for high granularity hodoscopes to be deployed in volcanic areas because of their small dimensions, robustness, and low power consumption \cite{Ambrosino2014}. SiPMs contain a dense array of small photon avalanche diodes operating in Geiger mode. When a photon interacts with a SiPM microcell, an avalanche process starts generating a photocurrent flowing through a quenching resistor, which causes that the diode bias drops below the breakdown value preventing further Geiger-mode avalanches. The electrical pulses generated by the SiPM are directly related to the number of incident photons. The main drawback of SiPMs is that performance parameters like gain, photodetection efficiency, and breakdown voltage are susceptible to temperature variations. Thermo-electric cells can control the SiPM temperature, but this methodology carries an increase in the power consumption \cite{Ambrosino2014}, which reduces the powering efficiency of autonomous muon telescopes. 

The Muon Telescope (MuTe) is a hybrid detector composed of a hodoscope and a Water Cherenkov Detector (WCD) which will be installed in one of the most dangerous volcanoes in Colombia, the Cerro Machin, located at 2750 m.a.s.l. on the Cordillera Central near to the municipality of Cajamarca \cite{AsoreyEtal2017B}. The MuTe hodoscope consists of two scintillation panels each of 30 $\times$ 30 strips $120$\,cm length, and 4 cm width. Each strip has a $1.8$\,mm hole for a multi-cladding wavelength shifting (WLS) fiber (Saint-Gobain BCF-92)  with $1.2$ mm diameter, an absorption peak at $410$\,nm and an emission peak of $492$\,nm \cite{SaintGobain2018}. The WLS fiber is coupled, with a silicon photomultiplier (SiPM, Hamamatsu S13360-1350CS) \cite{Hamamatsu2018}. The SiPM has a photosensitive area of $1.3 \times 1.3$\,mm$^2$, $667$ pixels, a fill factor of $74\%$, a gain from $10^5$ to $10^6$ and a photon-detection efficiency of $40 \%$ at $450$\,nm \cite{pena2019calibration, PenaRodriguezEtal2018, VsquezRamrez2020}.


This paper shows the characterization of the SiPM breakdown voltage, gain, dark count, crosstalk, and afterpulsing depending on temperature and over-voltage. In section \ref{sec:exp} we describe the experimental setup and the data acquisition system for the SiPM parameter measurements. The breakdown voltage, gain, and noise characterization results are described in section \ref{sec:res}. Section \ref{sec:obs} presents the temperature conditions at the Cerro Mach\'in volcano and their affectation on the MuTe mechanical structure, and section \ref{sec:temp} exhibits the dependence between the flux and the temperature of the MuTe tracking system in on-field conditions. Conclusions and remarks are summarized in section \ref{sec:conc}.

\section{Experimental setup}
\label{sec:exp}

\subsection{Breakdown voltage}

\begin{figure}[htbp]
\centering 
\includegraphics[width=.5\textwidth]{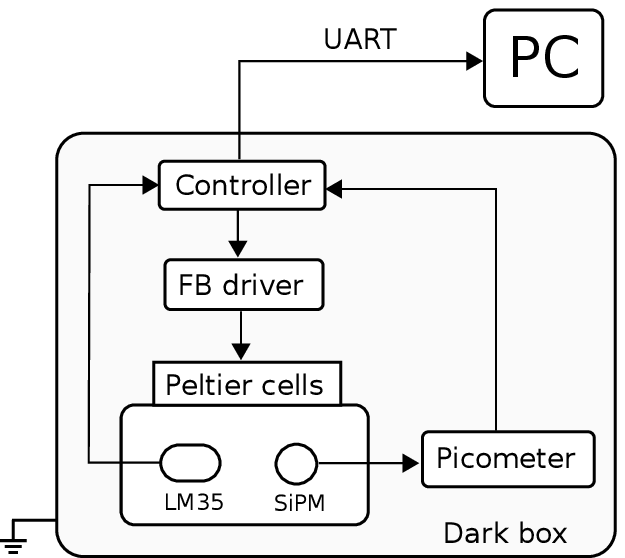}
\caption{\label{fig:setup1} Experimental setup for measuring the SiPM dark current in darkness conditions. The SiPM is positioned in the aluminium holder inside the dark box. The holder temperature is controlled via Peltier cells by means a PID controller implemented in a microcontroller Atmega328.}
\end{figure}

The first experimental setup measures the SiPM dark current depending on temperature and bias voltage. A sketch of the setup is shown in figure \ref{fig:setup1}. The SiPM is placed on an isolated aluminium holder whose temperature is controlled by two Peltier cells (TEC1-12706 from Hebei I.T) and measured using an LM35 sensor. A proportional-integral-derivative (PID) control (implemented in a microcontroller Atmega328) generates two pulse-width-modulate signals whose duty cycle depends on the control error. The error is defined as the difference between the measured temperature and the pre-established set-point. The control signals drive the direction (cooling or heating) and amplitude (fast or slow) of the current flowing through the Peltier cells using an H-bridge with an optically coupled isolator circuit. 


A C11204 power module biases the SiPM S13360-1350CS covering a voltage range from $40$\,V to $60$\,V. The dark current is measured by a $2$\,nA accuracy picoammeter. The SiPM bias voltage and temperature are recorded individually by a 10-bit analog to digital converter (ADC) with a sampling rate of $1$\,Hz. All the setup components are placed inside a grounded dark box to avoid external light contamination and electromagnetic interference.

\subsection{Gain and noise}

In the second experimental setup, we estimate the SiPM gain and noise at several temperatures and over-voltages after stimulating with pulsed light. The light source must fulfill two features: a wavelength matching the SiPM spectral sensitivity and a pulse width of the order of a few ns \cite{Georgiev2016, Eigen2019}.

\begin{figure}[htbp]
\centering 
\includegraphics[width=.6\textwidth]{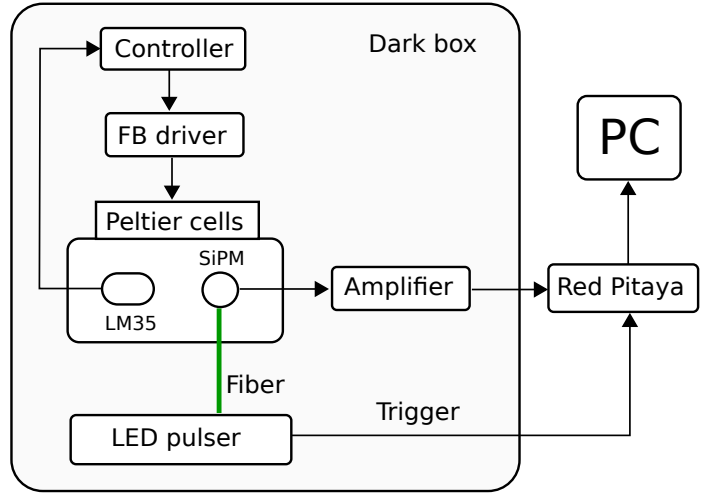}
\caption{\label{fig:setup2} Experimental setup for measuring the gain and noise of the SiPM S13360-1350CS under stimulated conditions. The SiPM is stimulated by a $480$\,nm pulsed light of $\sim10$\,ns width at $500$\,Hz. The SiPM signal is digitized by the Red Pitaya at 14-bit/$125$\,MHz.}
\end{figure}

The light pulser generates an ultra-short ($< 10$\,ns) $480$\,nm light pulse with a frequency of $500$\,Hz. A $50$\,cm WLS fiber (Saint-Gobain BCF-92) transports the light towards the SiPM, at the same time, a square signal triggers the DAQ system. The signals generated by the SiPM are amplified 94 times using a low noise current feedback operational amplifier (OPA691 from Texas Instruments) and digitized by a Red Pitaya ADC channel with a sampling frequency of $125$\,MHz and 14-bit resolution. A sketch of the setup is shown in figure \ref{fig:setup2}.




\section{SiPM calibration}
\label{sec:res}

\subsection{Breakdown voltage}
\begin{figure}[htbp]
\centering 
\includegraphics[width=.55\textwidth]{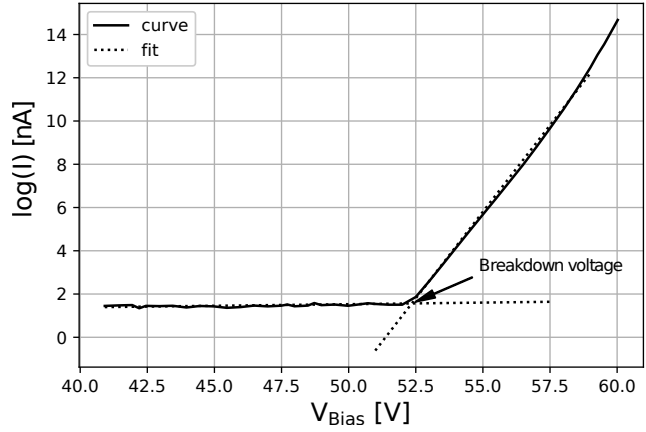}
\caption{Breakdown voltage value found using the tangent method for the IV curve of SiPM S13360-1350CS operating at $25^{\circ}$C. The V$_{br}$ ($52.3$\,V) is located at the intersection between the fit and the base line.}
\label{fig:Vbr} 
\end{figure}

The breakdown voltage (V$_{br}$) is the point where the SiPM enters in Geiger mode. Such a point can be established using several methods \cite{Nagy2017}. In this case, we use the tangent method which consists of finding the interception between a tangent line fitted to the IV (dark current vs bias voltage) curve and the baseline. In figure \ref{fig:Vbr} we show the SiPM IV curve at $25^{\circ}$C where the V$_{br}$ was found $\sim52.3$\,V.

\begin{figure}[htbp]
\centering 
\includegraphics[width=.47\textwidth]{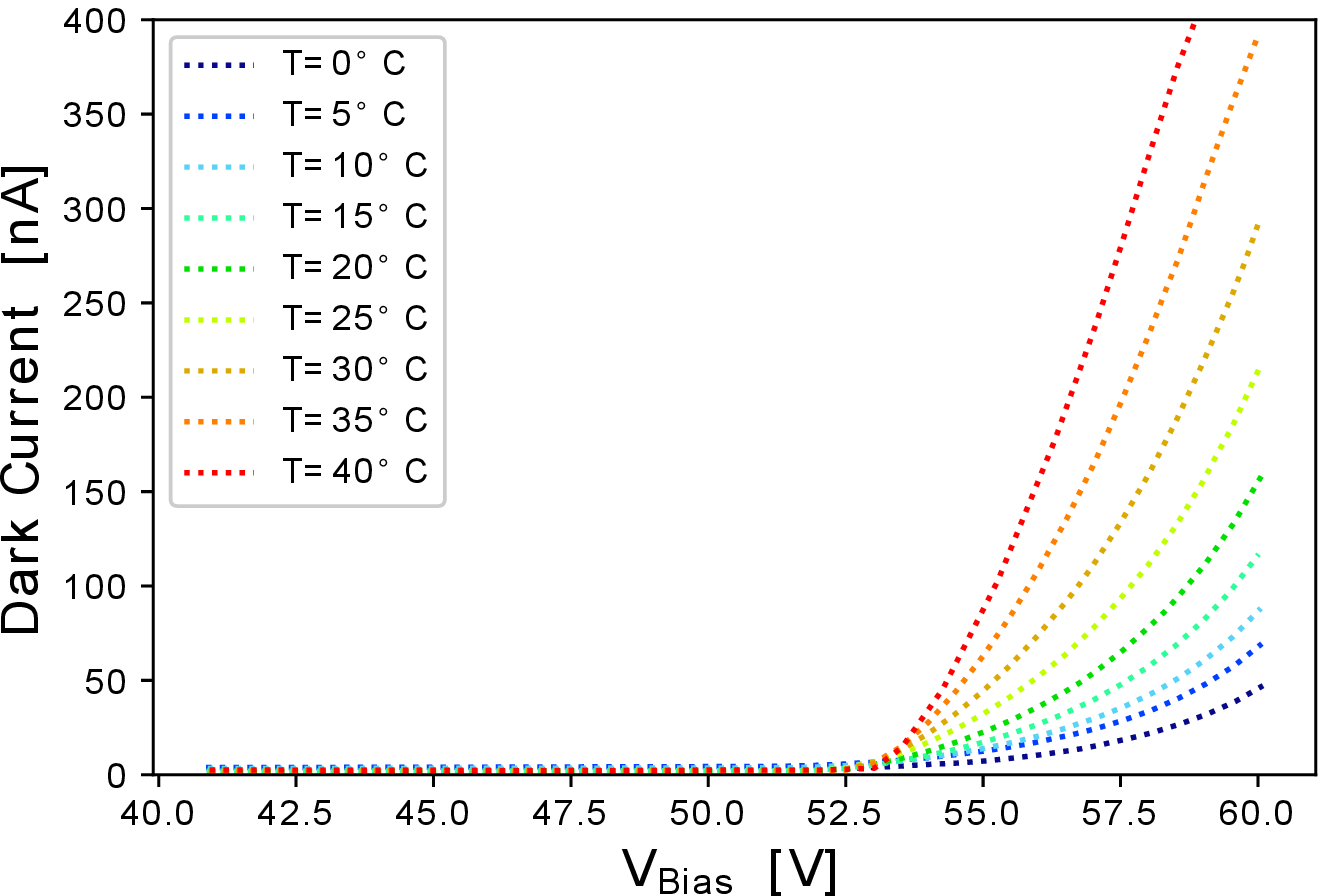}
\qquad
\includegraphics[width=.47\textwidth]{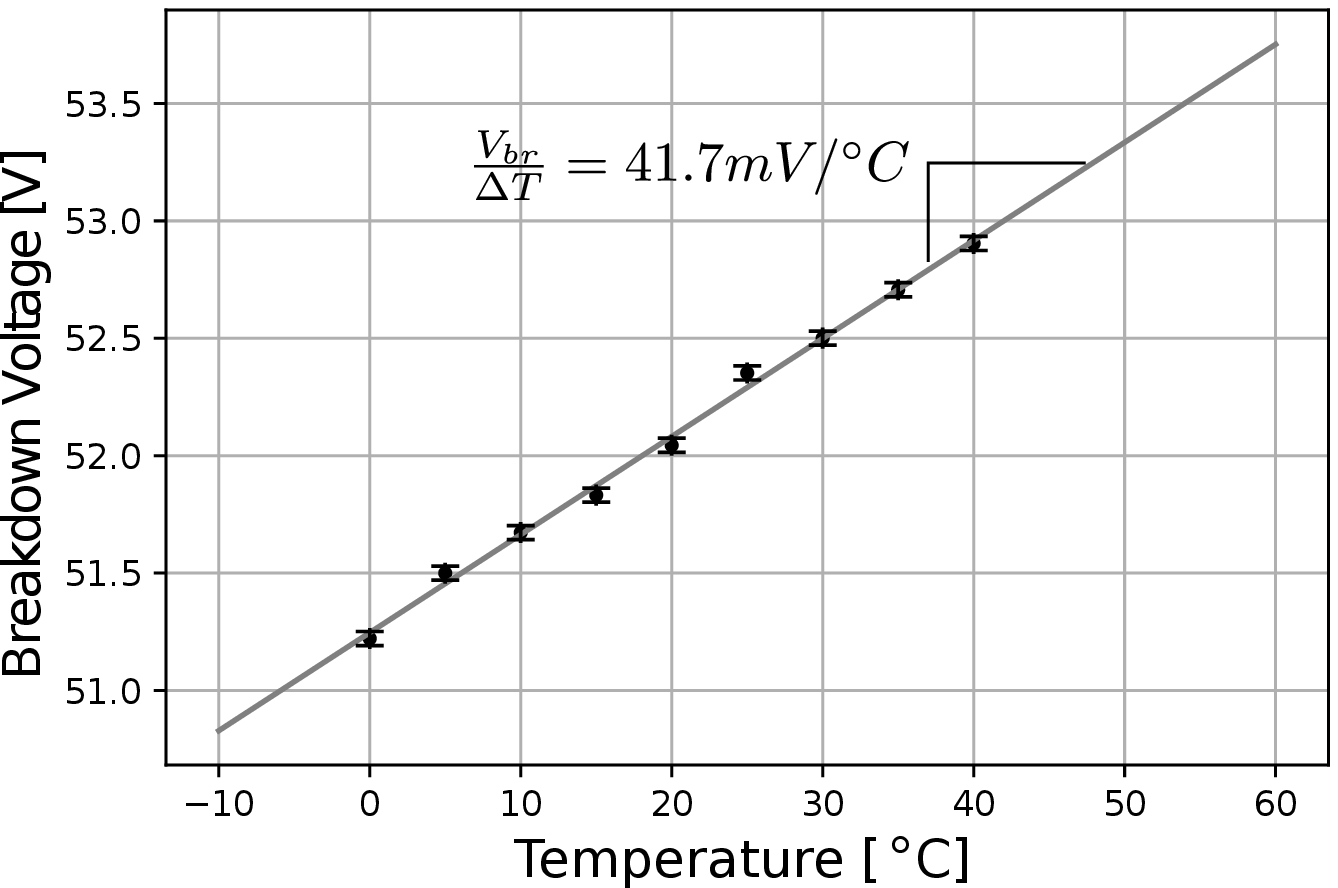}
\caption{\label{fig:IVcurve} Temperature dependence of the breakdown voltage for the SiPM S13360-1350CS from Hamamatsu. (Left): IV curves ranging from 0$^{\circ}$C to 40$^{\circ}$C. (Right): $V_{br}$ variance ratio depending on the temperature.}
\end{figure}

We measured the IV curves from $40$\,V to $60$\,V for temperatures between $0^{\circ}$C and $40^{\circ}$C with $5^{\circ}$C step as shown in figure  \ref{fig:IVcurve} (left-panel). In the Geiger region, the IV slope increases with the temperature reaching a dark current above $400$\,nA at $40^{\circ}$C. The breakdown voltage has a linear relation with temperature decreasing with a ratio of $41.7$\,mV/$^{\circ}$C as is shown in figure \ref{fig:IVcurve} (right-panel). In on-field applications, an adaptive bias voltage to compensate for temperature changes in the SiPM could be taken into account to guarantee a stable gain and low noise levels \cite{Eigen2019}.

\subsection{Charge spectrum and gain}

\begin{figure}[htbp]
\centering 
\includegraphics[width=.65\textwidth]{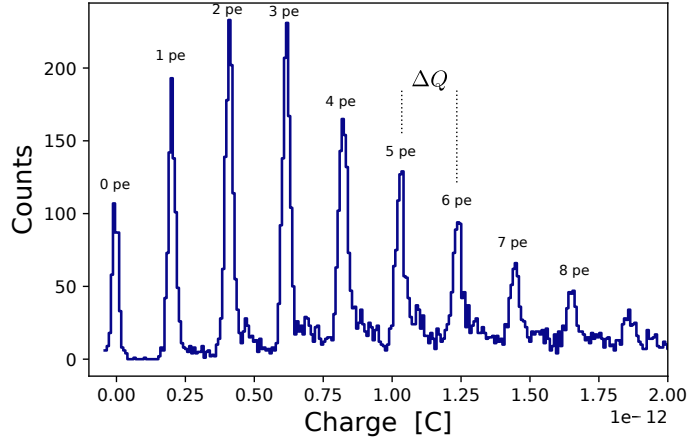}
\caption{\label{fig:charge} Charge spectrum of the SiPM S13360-1350CS operating at $56$\,V/$25^{\circ}$C. The first peak is the pedestal and the following represent the photoelectron equivalents. The inter-peak charge $\Delta Q$ determines the SiPM gain.}
\end{figure}

The gain of a SiPM microcell is defined as the ratio of the output charge to the charge on an electron $e$ \cite{Acerbi2019}. The output charge can be calculated as,

\begin{equation}
\label{eq:charge}
Q = \frac{Q_{ADC}V_{ADC}\Delta_T}{RG_a}
\end{equation}
where $Q_{ADC}$ is the digitized area under pulse, $V_{ADC}$ is the equivalent voltage for one ADC unit, $\Delta_T$ is the digitization time step, $R$ is the input resistor and $G_a$ the gain of the electronics front-end. In figure \ref{fig:charge} the charge spectrum of the SiPM operating at $56$\,V and $25^{\circ}$C is shown.

The separation between two adjacent peaks $\Delta Q$ in the charge histogram corresponds to the charge from a single Geiger discharge. This can be used to accurately calculate the gain $G$ as follows,

\begin{equation}
\label{eq:gain}
G = \frac{\Delta Q}{e}
\end{equation}

\begin{figure}[htbp]
\centering 
\includegraphics[width=.49\textwidth]{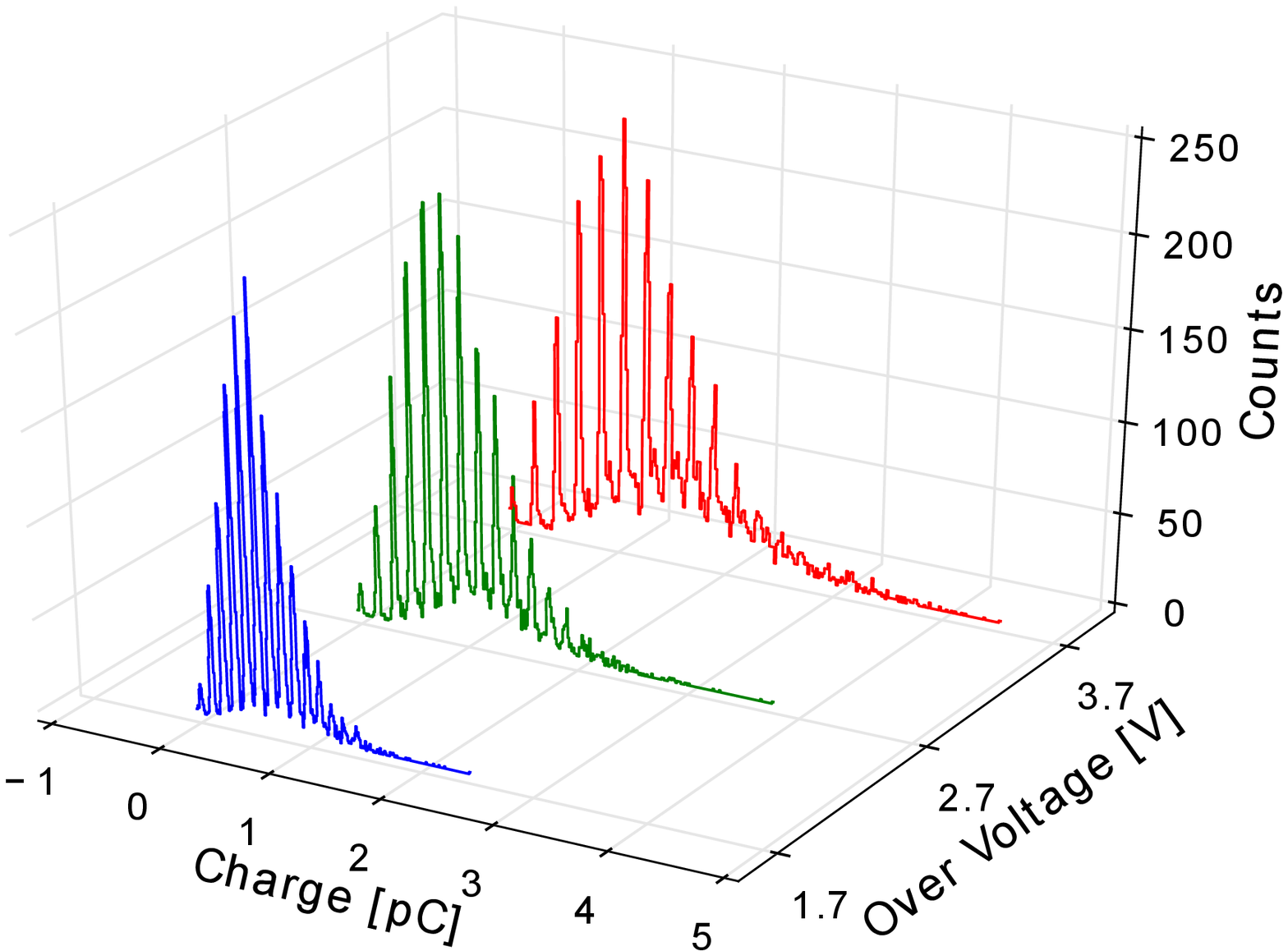}
\quad
\includegraphics[width=.47\textwidth]{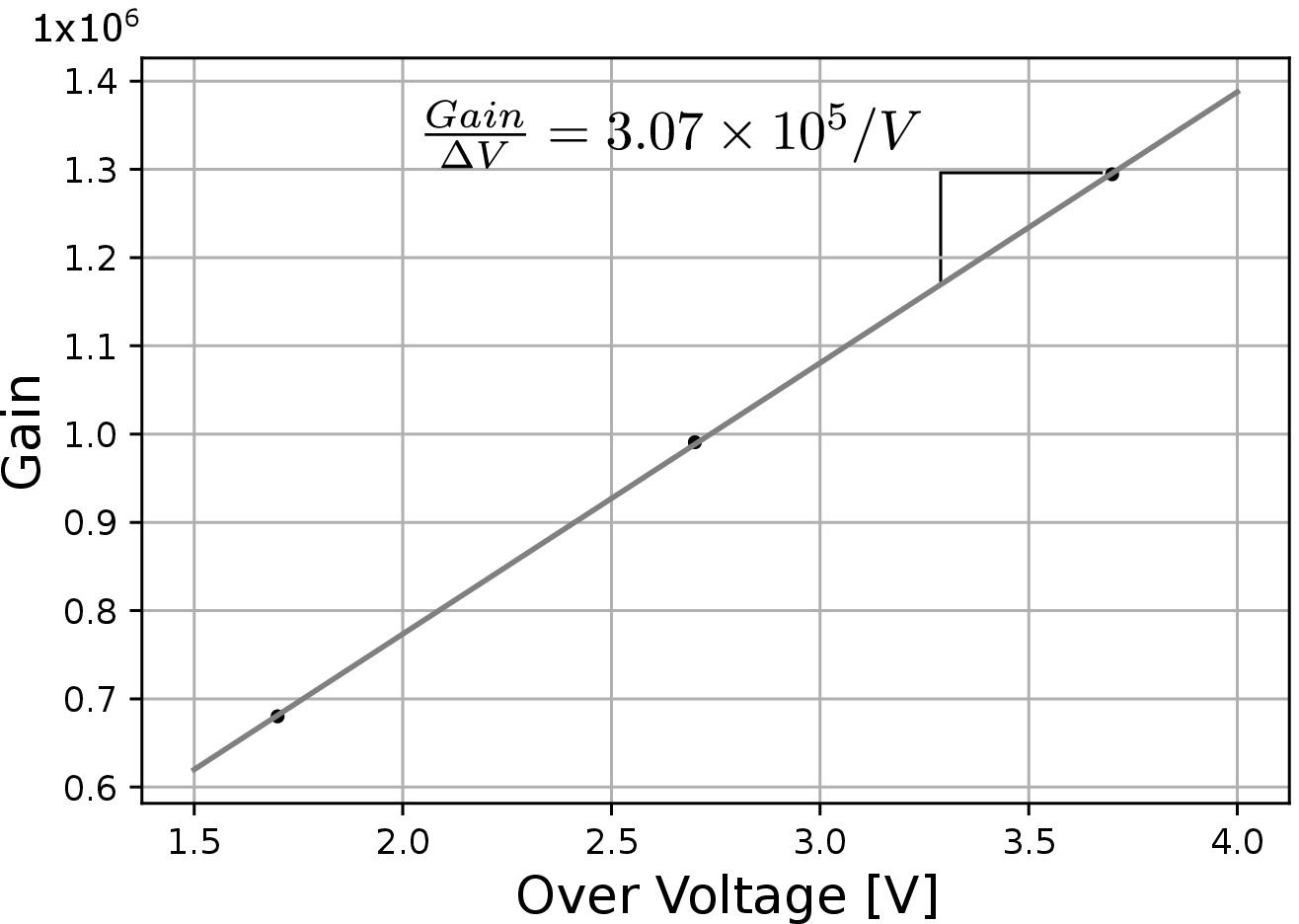}
\caption{\label{fig:gain} (Left): Charge spectrum for $\Delta V=1.7$\,V (blue), $2.7$\,V (green), and $3.7$\,V (red). (Right): Gain variation ratio depending on the over-voltage.}
\end{figure}

The SiPM gain depends on the bias voltage ($V_{bias} = V_{br} + \Delta$V), the higher the bias voltage the higher the gain. To estimate the gain dependence on the over-voltage ($\Delta V$) in the SiPM S13360-1350CS we measured three charge spectra for $\Delta$V = $1.7$\,V, $2.7$\,V, and $3.7$\,V at $25^{\circ}$C. Figure \ref{fig:gain} shows the charge spectra (left-panel) and the estimated gain (right-panel) for these cases.

The separation between charge peaks grows as the over-voltage increases --indicating a gain increment. The gain change ratio was estimated $\sim3.07\times10^5$/V, i.e., for $\Delta$V = $1.7$\,V ($V_{bias} = 53$\,V) the gain is roughly $0.7\times10^6$ and $\Delta$V = $3.7$\,V ($V_{bias} = 56$\,V) the gain is $1.3\times10^6$.

\subsection{Photoelectron spectrum}

The output pulse amplitude from SiPMs is proportional to the number of incident photons based on the fact they are made of an array of APDs connected in parallel. The photoelectron spectrum determines the equivalent value (voltage or current) of a photon interacting with the active area of the SiPM. This value establishes the threshold for measuring dark count rate (DCR), crosstalk and afterpulsing noise.


\begin{figure}[htbp]
\centering 
\includegraphics[width=.48\textwidth]{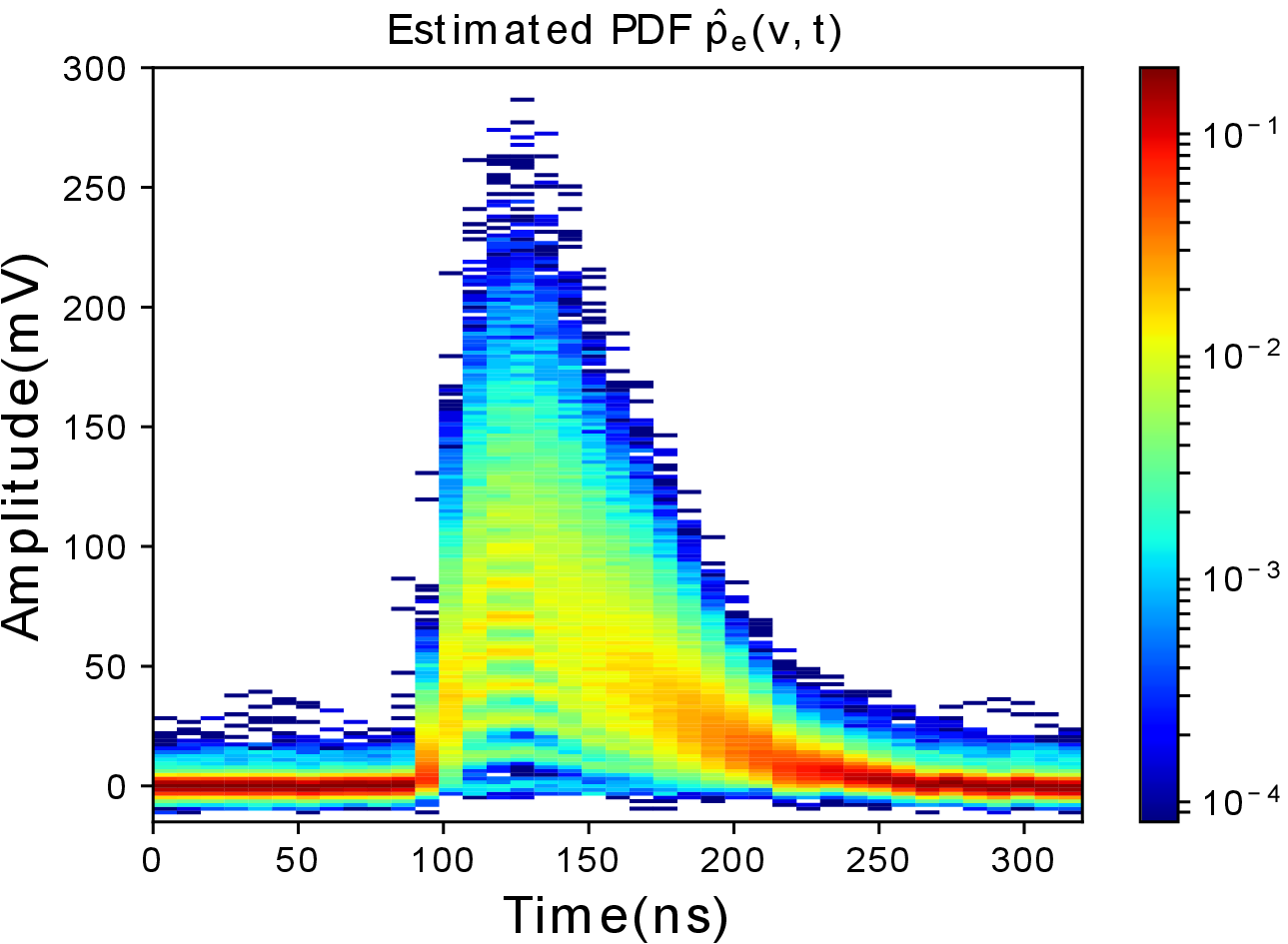}
\includegraphics[width=.49\textwidth]{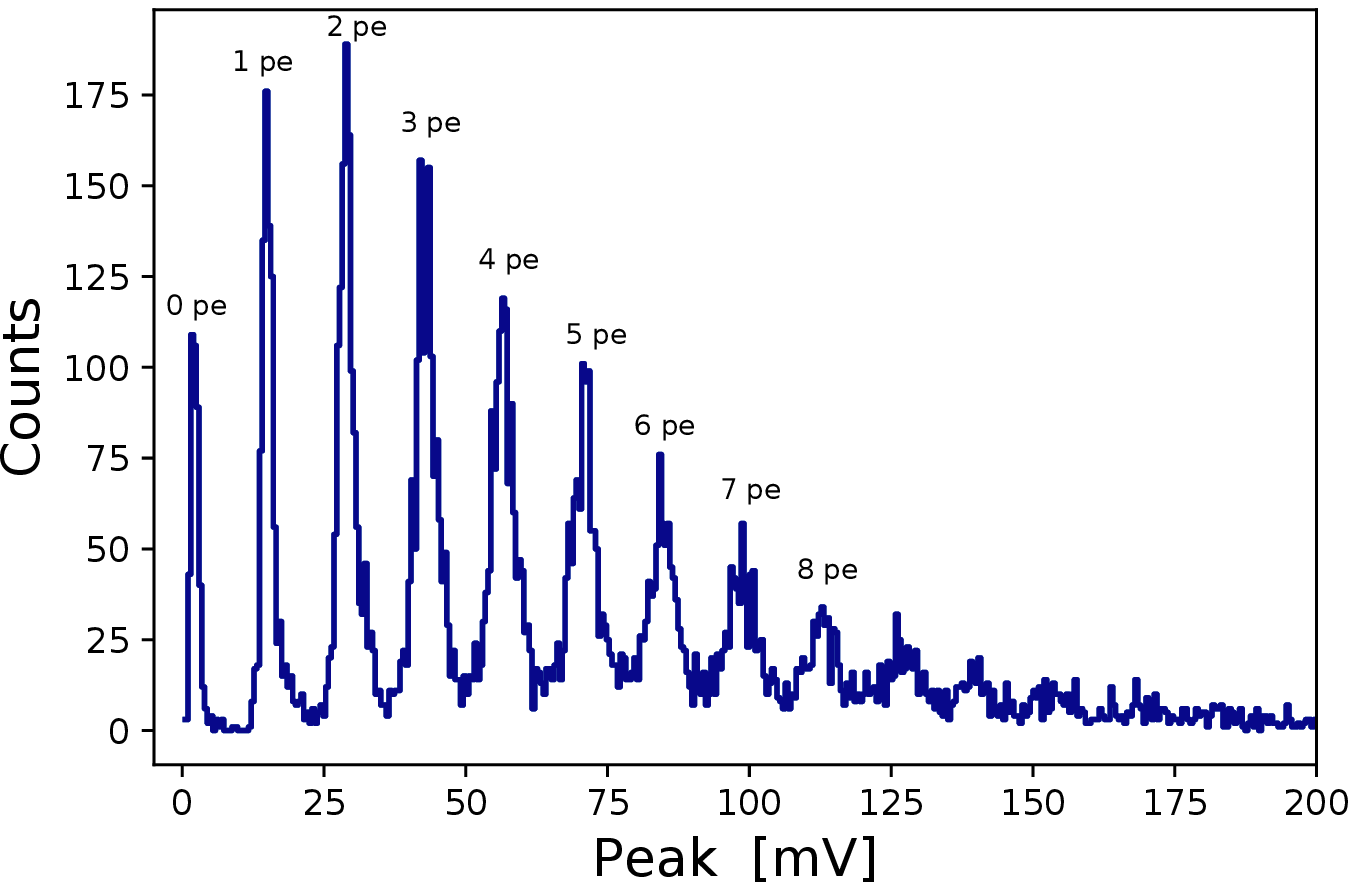}
\caption{\label{fig:peak} (Left): Waveform of the Hamamatsu S13360-1350CS under stimulation. Photo-electron spectrum resulting from integrating the area under pulse over a time window of $300$\,ns.}
\end{figure}

In figure \ref{fig:peak} the persistence histogram (left-panel) of the pulse shape and the peak histogram (right-panel) for $10^4$ pulses at $56$\,V/$25^{\circ}$C are shown.

The histograms reveal that pulses of $1$\,pe and $2$\,pe have more probability of occurrence than others. These pulses are mainly generated by the SiPM noise. The resulting equivalent voltage for $1$\,pe is $\sim13.5$\,mV, therefore the threshold for measuring the SiPM DCR must be set below $13.5$\,mV and for the cross-talk below $27$\,mV.

\subsection{Noise}

SiPMs are affected by correlated noise (crosstalk and afterpulsing) and non-correlated noise (DCR) \cite{Baudis2018}. These noise sources impose the lower measurement limit in SiPM based experiments. We performed a noise analysis of the MuTe SiPMs taking into account its temperature and over-voltage dependency. We also established the minimum pe threshold above which the noise is negligible.

\subsubsection{Dark count rate}
The main source of noise in SiPMs is the DCR. It appears as a consequence of avalanches processes fired by electrons thermally generated in the silicon crystal. Signals generated by thermal electrons and single-photons are identical. The DCR is measured under dark conditions by counting events above a $0.5$\,pe threshold.

\begin{figure}[htbp]
\centering 
\includegraphics[width=.55\textwidth]{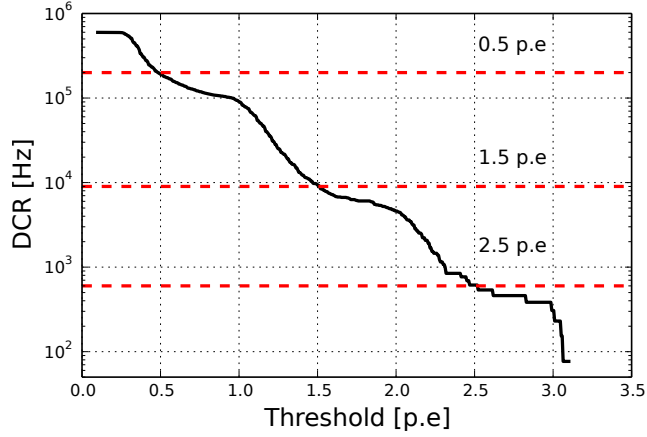}
\caption{\label{fig:DCR} Dark count rate as a function of the detection threshold. The curve shape presents three breaks at $1$\,pe, $2$\,pe and $3$\,pe because of the discretization effect on the pulse amplitude.}
\end{figure}

The DCR is calculated as follows,
\begin{equation}
    DCR = \frac{N_{1pe}^B}{T^B N_p}
\label{eq:DCR}
\end{equation}
where $N_{1pe}^B$ is the number of events above 0.5\,pe in the time window $T^B$ (before stimulation) and $N_p$ is the total number of recorded events.

We measured the DCR for different thresholds spanning from $0.1$\,pe to $3.1$\,pe at $56$\, V/$25^{\circ}$C as shown in figure \ref{fig:DCR}. The resulting curve has a stepped shape because of the amplitude discretization of the SiPM pulses. At $0.5$\,pe the DCR is $\sim2\times 10^5$ Hz corresponding with the expected value provided by the SiPM S13360-1350CS datasheet which range between $0.9\times 10^5$ Hz and $2.7\times 10^5$ Hz.

The DCR drastically decreases while the measurement threshold increases. We found a DCR of $9\times 10^3$\,Hz at $1.5$\,pe and $6\times 10^2$\,Hz at $2.5$\,pe At $3.5$\,pe the DCR is expected to be negligible ($< 10$\,Hz).

\begin{figure}[htbp]
\centering 
\includegraphics[width=.48\textwidth]{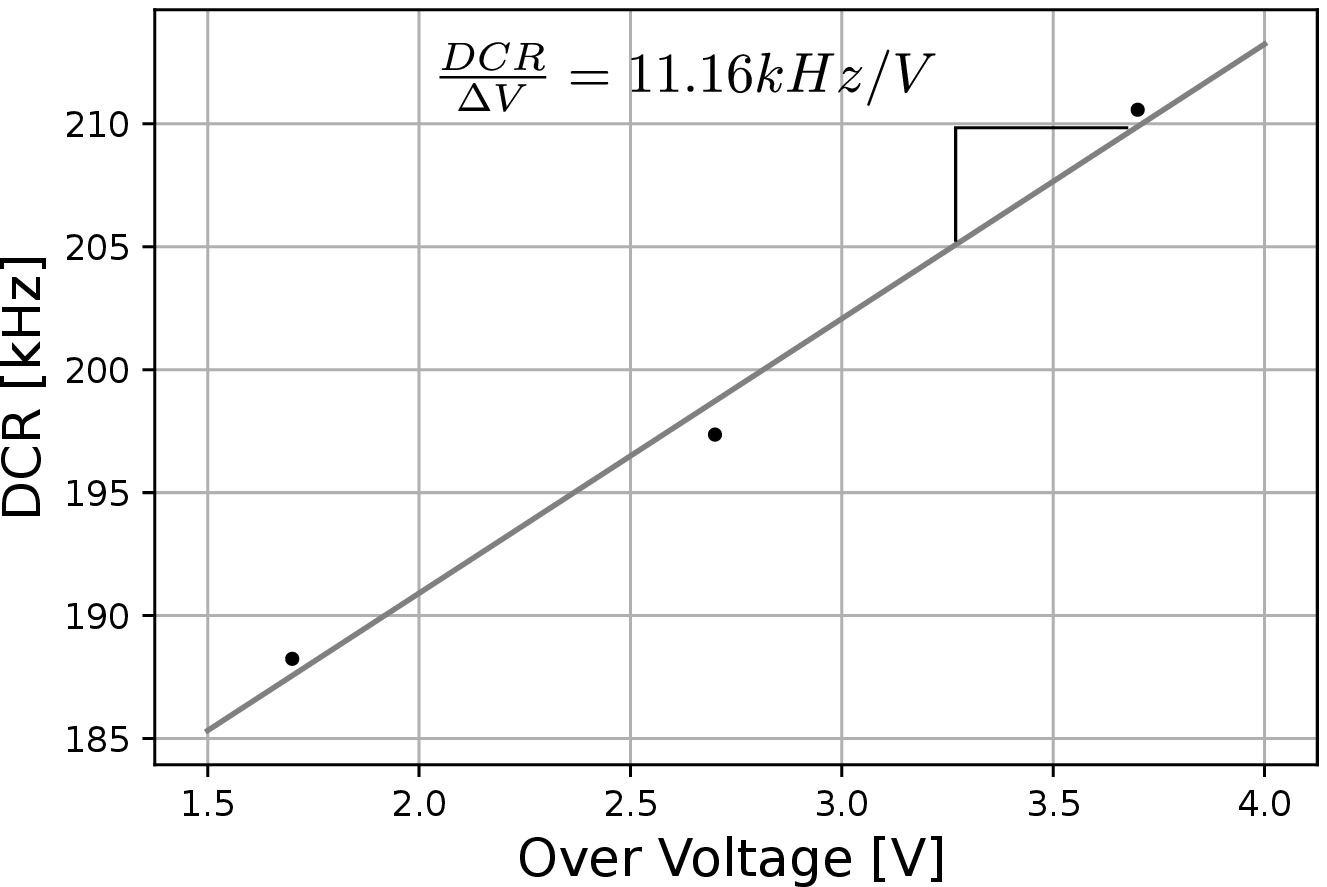}
\quad
\includegraphics[width=.48\textwidth]{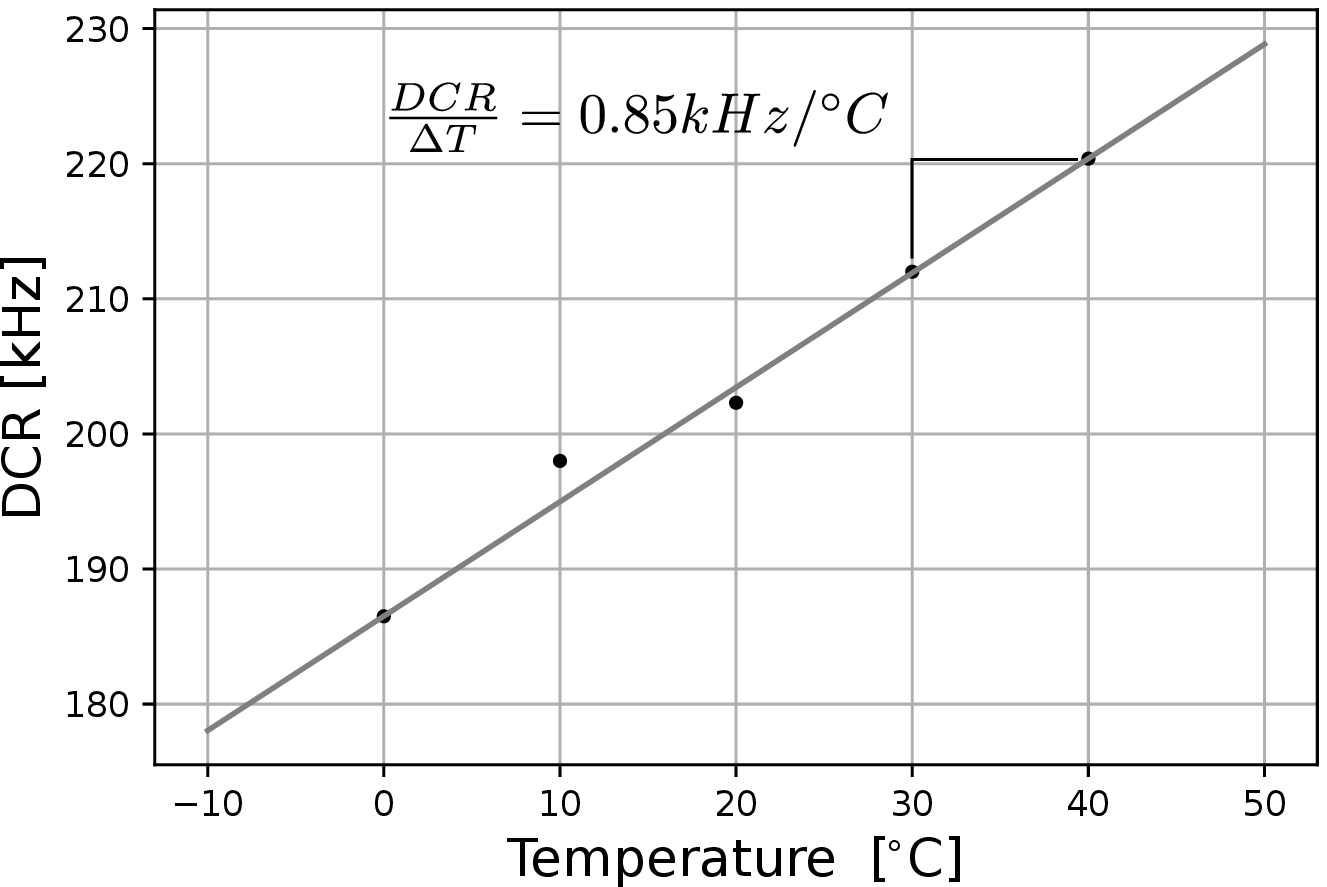}
\caption{\label{fig:DCRover} (Left): Dark count rate as a function of the over-voltage spanning from ($1.7$\,V to $3.7$\,V) at $25^{\circ}$C. (Right): Dark count rate as a function of temperature spanning from $0^{\circ}$C to $40^{\circ}$C at $56$\,V. The variation ratio is $0.85$\,kHz/$^{\circ}$C.}
\end{figure}

To characterize the DCR as a function of the over-voltage, we carried out DCR measurements for three cases ($1.7$\,V, $2.7$\,V and $3.7$\,V) at $25^{\circ}$C. Figure \ref{fig:DCRover} (left-panel) shows that the DCR increases with a slope $\sim11.16$\,kHz/V. 

The DCR correlation with the SiPM temperature was also evaluated. We estimated a ratio $0.85$\,kHz/$^{\circ}$C after analyzing DRC measurements from $0^{\circ}$C to $40^{\circ}$C at $56$\,V as shown figure \ref{fig:DCRover} (right-panel).

\subsubsection{Afterpulsing and crosstalk}

Afterpulsing is generated by trapped electrons in silicon impurities during an avalanche process. These electrons are released few nanoseconds later creating new avalanches --consecutive pulses \cite{Xu2017}. The amplitude of afterpulses increases with the retention time of the trapped electron. 



The afterpulsing probability $P_{AP}$ is calculated as follows
\begin{equation}
    P_{AP} = \frac{N_{1pe}^A-N_{1pe}^B}{N_p}\times 100
\label{eq:AP}
\end{equation}
where $N_{1pe}^A$ is the number of events above $0.5$\,pe in the time window $T^A$ (after stimulation). 

Crosstalk occurs when charge carriers (inside the avalanche) emit photons that interact with neighboring cells. Such interactions trigger secondary avalanches in these cells with amplitudes of $2$\,pe or $3$\,pe.



The crosstalk probability \cite{Ramilli2008} is defined as
\begin{equation}
    P_{CT} =\frac{N_{2pe}^B}{N_{1pe}^B}\times 100
\label{eq:CT}
\end{equation}
where $N_{2pe}^B$ is the number of events with amplitude above $1.5$\,pe before stimulation.

\begin{figure}[htbp]
\centering 
\includegraphics[width=.48\textwidth]{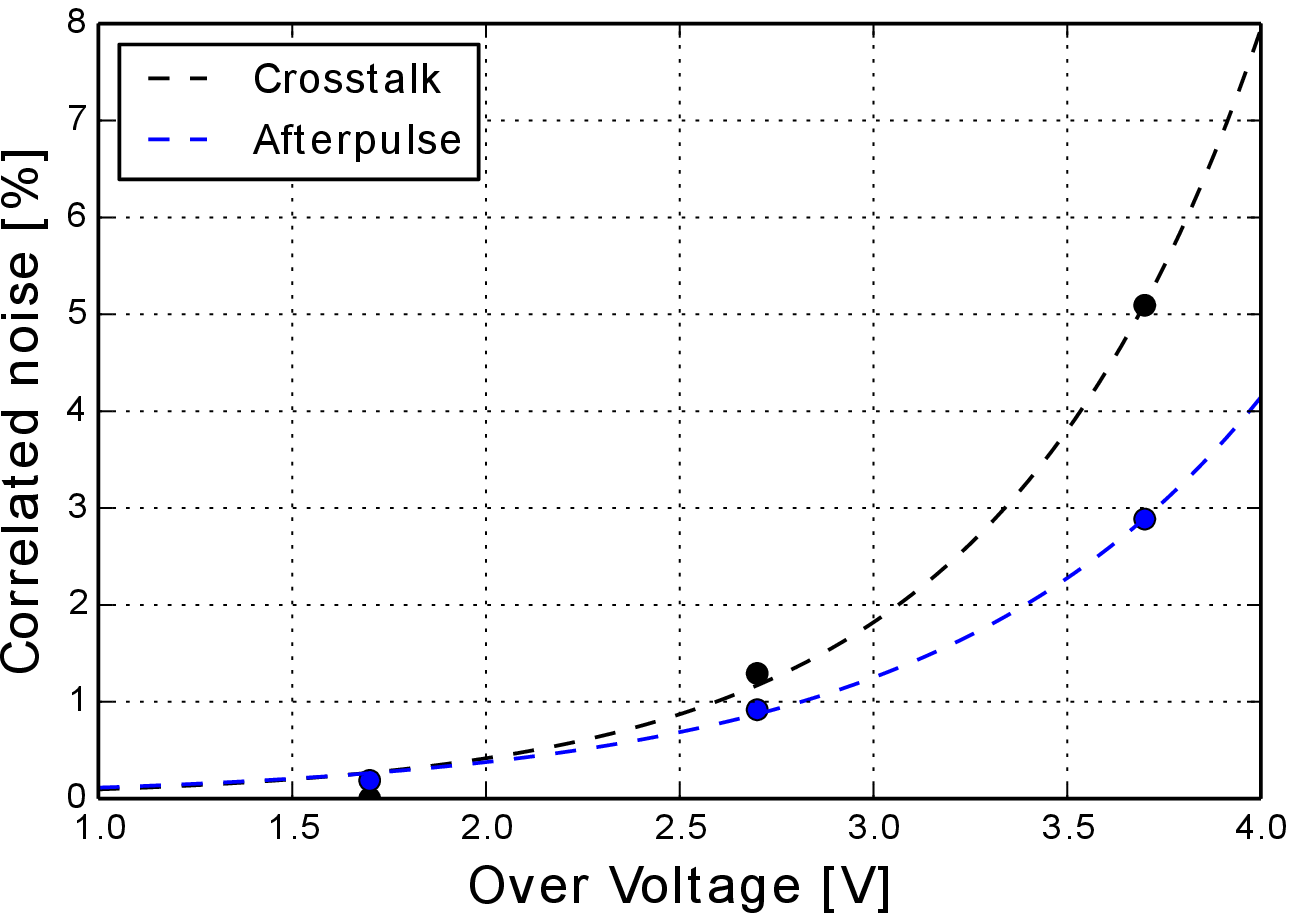}
\quad
\includegraphics[width=.48\textwidth]{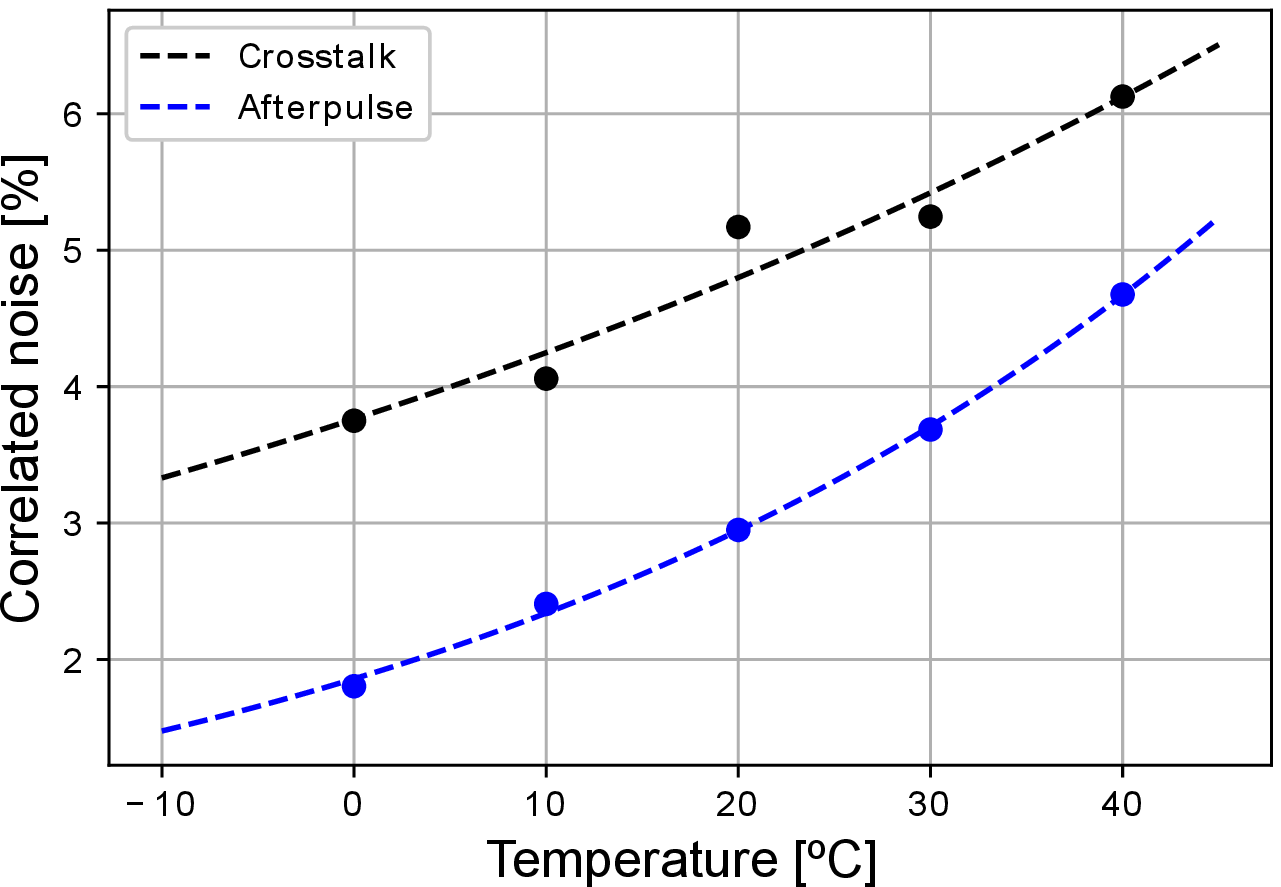}
\caption{MuTe-SiPM crosstalk (black line) and afterpulsing (blue line) depending on its over-voltage (left) and temperature (right).}
\label{fig:after_cross}
\end{figure}

Figure \ref{fig:after_cross} (right-panel) shows the afterpulsing and crosstalk versus the SiPM over-voltage. Both increase exponentially with the over-voltage, being the crosstalk greater than afterpulsing. The afterpulsing reaches $3\%$ at $56$\,V/$25^{\circ}$C while the crosstalk $5\%$.

The correlated noise dependency on temperature was analyzed by performing afterpulsing and crosstalk measurements from $0^{\circ}$C to $40^{\circ}$C at $56$\,V. The results are displayed on Figure \ref{fig:after_cross} (Left). At 0$^{\circ}$C the afterpulsing probability is below $2\%$ and the crosstalk below $4\%$. The afterpulsing increases faster than crosstalk with the temperature, rising up almost $5\%$ at $40^{\circ}$C while crosstalk reaches $6\%$.

To reduce the noise caused by dark count, crosstalk, and afterpulsing, we concluded that the minimum discrimination threshold for the scintillator hodoscope of MuTe must be above $5$\,pe. The breakdown voltage shifting due to temperature variations will cause a modulation of the detection rate. This can be solved using closed-loop control of the SiPMs bias voltage or corrected in the offline data analysis.

\section{Operating temperature conditions of the MuTe}
\label{sec:obs}

\subsection{Weather at the Cerro-Machin Volcano}

The Cerro-Mach\'in volcano has typical weather conditions of the Andean mountains in Colombia. According to the Colombian Hydrology, Meteorology and Environmental Studies Institute (IDEAM), at the Cerro-Mac\'in the average temperature is $16^{\circ}$C, the relative humidity $85\%$, and the maximum wind speed $30$\,m/s. During the rainy season, the temperature drops to $0^{\circ}$C, and during the dry season, it rises to $25^{\circ}$C. The rainy season comes from April to May and October to November, and the dry season is usually from December to January and July to August. The day-night temperature gradient at the Cerro-Mach\'in volcano is around $10^{\circ}$C along the dry and rainy seasons.

\subsection{Heat transfer in the MuTe structure}

We computed a thermal analysis of the MuTe mechanical structure using the \textsc{Solidworks CAD Software}. The heat sources were: the environmental temperature (16$^{\circ}$C), solar radiation (4.5 kWh m$^{-1}$day$^{-1}$), cooling by wind (30 m/s), and heating by electronics power consumption (12.5 W). We also input thermal features of the metallic chassis supporting the WCD and the hodoscope \cite{penarodriguez2020}.

\begin{figure}[htbp]
\centering 
\includegraphics[width=.7\textwidth]{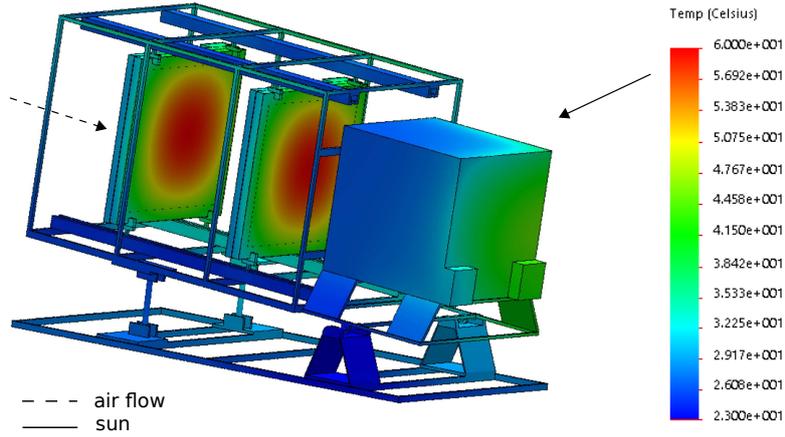}
\caption{Heat distribution of the MuTe structure under environmental conditions at the Cerro-Machin volcano. The solid-arrow represents the incident solar radiation while the dashed-arrow indicates the wind direction. The maximum temperature at the center of the scintillation panels reaches $60^{\circ}$C. }
\label{fig:detec} 
\end{figure}

Figure \ref{fig:detec} displays the temperature distribution on the MuTe structure resulting from the thermal simulation. The direct incidence of the solar radiation (solid arrow) causes a maximum temperature of $60^{\circ}$C in the middle of the scintillation panels, but this drops to $26^{\circ}$C due to the convection created by the frontal wind (dashed arrow). The water volume inside the WCD dissipates the heat of the stainless steel cube. The maximum temperature on the WCD is $\sim40^{\circ}$C.

\section{Temperature influence on the MuTe-SiPMs}
\label{sec:temp}

In this section, we analyze how temperature affects the SiPM parameters under real observation conditions. This procedure uses the characterization ratios presented above and temperature measurements.

\begin{figure}[htbp]
\centering 
\includegraphics[width=.4\textwidth]{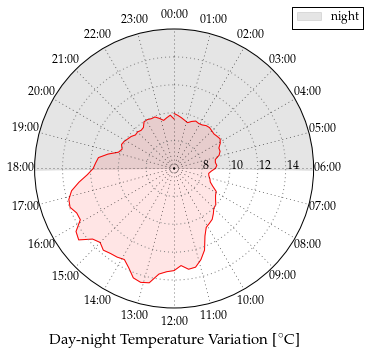}
\caption{\label{fig:temp_mach} Day-night temperature cycle at the Cerro-Mach\'in volcano during the rainy season (November 22-23). The gray shadow indicates the night period starting at 18:00 and ending at 06:00. The minimum temperature ($\sim8.5^{\circ}$C) is recorded at 06:30 and the maximum ($\sim14.5^{\circ}$C) at 13:00. }
\end{figure}

We use temperature data recorded at the Cerro Machin volcano during the 2017 rainy season between November 22-23. The day-night temperature cycle stars/ends at the 00:00 hour with $\sim10^{\circ}$C. The temperature drops to a minimum value of $\sim8.5^{\circ}$C at morning (06:30) and rises to a maximum of $\sim14.5^{\circ}$C at day (13:00) as shown figure \ref{fig:temp_mach}. 

\begin{figure}[htbp]
\centering 
\includegraphics[width=.4\textwidth]{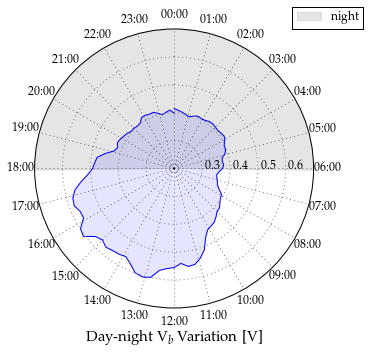}
\includegraphics[width=.4\textwidth]{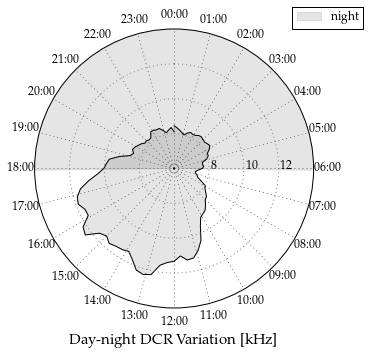}
\caption{\label{fig:behav} MuTe SiPM breakdown voltage and DCR variation as a function of typical temperature values at the Cerro-Mach\'in volcano.}
\end{figure}

The estimated SiPM breakdown voltage and DCR along the day-night cycle is presented in figure \ref{fig:behav}. The maximum temperature gradient is $\sim6.1^{\circ}$C which represents a breakdown voltage ($41.7$\,mV/$^{\circ}$C) deviation of $\pm126$\,mV from the nominal value ($53.2$\,V). This breakdown shift affects the SiPM gain ($3.07\times10^5$/V) causing a deviation $\sim0.8\times 10^5$.


As the temperature on the SiPM increases, the number of thermally generated electrons on the silicon material also increases. The DCR absolute variation is $\sim5.2$\,kHz. We can expect the DCR varies between $207$\,kHz and $212.2$\,kHz assuming the SiPM operates at $56$\,V where the nominal DCR rounds $210$\,kHz. 

\begin{figure}[htbp]
\centering 
\includegraphics[width=.4\textwidth]{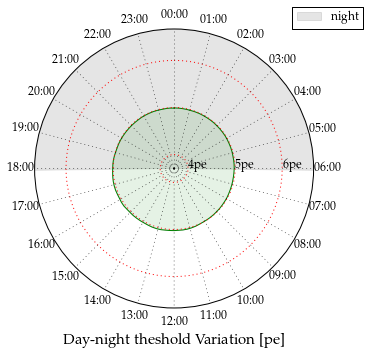}
\includegraphics[width=.4\textwidth]{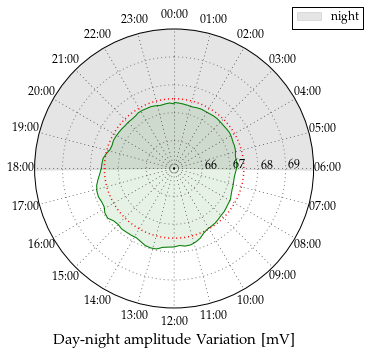}
\caption{\label{fig:thre} Photoelectron and pulse amplitude variation of the MuTe-SiPM for typical temperature values at the Cerro-Mach\'in volcano.}
\end{figure}

The maximum variance of the pulse amplitude is about $0.8$\,mV for a $\Delta T \sim 6.1^{\circ}$C which represents roughly $6\%$ of the voltage separation between two consecutive photoelectrons ($\sim13.5$\,mV at $56$\,V). The MuTe hodoscope discrimination threshold was set above $8$\,pe = 108.5\,mV to take out the noise contributions due to DCR, afterpulsing, and crosstalk. Figure \ref{fig:thre} shows the resulting variance around the threshold voltage (Right) and its respective photoelectron equivalent (Left).


\begin{figure}[htbp]
\centering 
\includegraphics[width=0.7\textwidth]{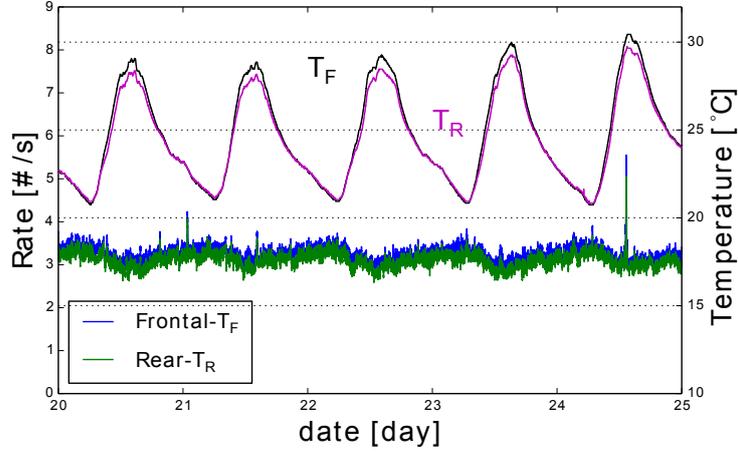}
\caption{\label{fig:temp} Hodoscope rate modulation depending on the environmental temperature of the MuTe recorded from 2019/12/20 to 2019/12/25. The green line displays the rear panel rate under temperature $T_R$ and the blue line the frontal panel rate under temperature $T_F$. }
\end{figure}

We analyzed 5 days of data from 2019/12/20 to 2019/12/25 to evaluate the MuTe-SiPMs behavior in on-field conditions. In Figure \ref{fig:temp} we show the temperature of the rear ($T_R$) and frontal ($T_F$) panels, as well as the in-coincidence detection rate. The MuTe was set pointing towards the horizon, with an angular aperture of 52$^{\circ}$, and an inter-panel separation of 2.5\,m.

The panel temperature oscillates from 20$^{\circ}$C to 30$^{\circ}$C representing a gradient of 10$^{\circ}$C. A 10$^{\circ}$C gradient represents a variation in the pulse amplitude around 14.8$\%$.
But this variation increases the breakdown voltage $\sim$417\,mV, reducing the overvoltage and the SiPM gain causing a reduction of the detected rate. The measured average flux is $\sim$3.1 events/s varying roughly 11.2$\%$ at the maximum and minimum temperature. The hodoscope recorded flux is corrected offline taking into account the estimated temperature dependence -0.057\,Hz/$^{\circ}$C \cite{penarodriguez2021}.


\section{Conclusions and Outlook}
\label{sec:conc}

We evaluated the SiPM S13360-1350CS from Hamamatsu to characterize the breakdown voltage, gain, and noise depending on the over-voltage and temperature. Temperature testes ranged from 0$^{\circ}$C to 40$^{\circ}$C covering the temperature spectrum of the observation site at Cerro Mach\'in Volcano, Colombia. The SiPM breakdown voltage variation ratio was about 41.7mV/$^{\circ}$C indicating a pulse amplitude shift of 14.8$\%$, which is not representative for jumping between photoelectron levels. We also estimated a gain increase ratio of about 3.07$\times 10^5$/V for over-voltage changes on the SiPM. 

In the noise characterization, we found that the dark count rate decreases by several magnitude orders (< 100 Hz) at a threshold above 3\,pe On the other hand, the DCR increases with a ratio of 11.16 kHz/V as a function of the SiPM over-voltage. This proposes a trade-off challenge because temperature increase generates a breakdown voltage increase but also a rising of the DCR. In the SiPM S13360-1350CS, the afterpulsing and crosstalk probabilities showed a non-linear growth with the temperature reaching up to 3$\%$ and 5$\%$ at an over-voltage of 3.7 V respectively. We recommend a discrimination threshold above 5\,pe to reduce drastically the correlated and non-correlated noise from MuTe SiPMs. 

In the on-field test, the hodoscope rate was modulated by the environmental temperature reaching a maximum deviation of 11.2$\%$ with respect to the average. The modulation was inversely correlated to the temperature (-0.057\,Hz/$^{\circ}$C) because of the breakdown voltage increase and the SiPM gain reduction.


\acknowledgments

The authors acknowledge the financial support of  Departamento Administrativo de Ciencia, Tecnolog\'ia e Innovaci\'on of Colombia (ColCiencias) under contract FP44842-082-2015 and to the Programa de Cooperaci\'on Nivel II (PCB-II) MINCYT-CONICET-COLCIENCIAS 2015, under project CO/15/02.


\bibliographystyle{unsrt}
\bibliography{MuTe.bib}

\end{document}